\documentstyle[epsf,epsfig,preprint,aps,amssymb]{revtex}
\draft \preprint{SNUTP 01/031}
\begin{document}
\title{\Large\bf
$Z_N$ orbifold compactifications in $AdS_6$ with Gauss-Bonnet term
}
\author{Jihn E. Kim\footnote{jekim@phyp.snu.ac.kr} and Hyun Min
Lee\footnote{minlee@phya.snu.ac.kr}}
\address{Department of Physics and Center for Theoretical
Physics, Seoul National University, Seoul 151-747, Korea}
\maketitle

\begin{abstract}
We present a general setup for junctions of semi-infinite 4-branes
in $AdS_6$ with the Gauss-Bonnet term. The 3-brane tension at the
junction of 4-branes can be nonzero. Using the brane junctions as
the origin of the $Z_N$ discrete rotation symmetry, we identify
3-brane tensions at three fixed points of the orbifold $T^2/Z_3$
in terms of the 4-brane tensions. As a result, the three 3-brane
tensions can be simultaneously positive, which enables us to
explain the mass hierarchy by taking one of two branes apart from
the hidden brane as the visible brane, and hence does not
introduce a severe cosmological problem.
\end{abstract}


\newpage

\section{Introduction}
The Higgs mass problem of the Standard Model (SM) includes ($i$)
the quadratic divergence of the Higgs mass squared and ($ii$) the
quartic divergence of vacuum energy density, which render power
law sensitivities to the unknown ultraviolet physics.
Supersymmetry in 4 dimensional(4D) spacetime has been thought as a
possible solution of this hierarchy problem. Recently, however,
extra dimensional scenarios have been suggested as alternative
solutions of the gauge hierarchy problem with the idea that the SM
particles are confined to a 3-brane embedded in a higher
dimensional spacetime\cite{ADD,rs1}.

In particular, the Randall-Sundrum (RS) I model\cite{rs1} with two
3-branes embedded in a slice of $AdS_5$, the gauge hierarchy
problem can be solved by taking the brane at the weak scale(TeV
brane) as the visible brane through the warp factor decreasing
away from the hidden brane(Planck brane). As a result, the
stability of the gauge hierarchy becomes the mechanism of radius
stabilization at appropriate separation of the Planck and TeV
branes\cite{stab}. On the other hand, the RS II model\cite{rs2}
cannot serve as a solution of the gauge hierarchy problem, but
presumably renders a more important paradigm toward an alternative
to compactification with a single 3-brane embedded in noncompact
$AdS_5$. It is based on the fact that the linearized Einstein
gravity can be reproduced on the brane even for the noncompact
extra dimension of the RS II model\cite{gravity}. Even with these
interesting results of the RS models, the notorious cosmological
constant problem is not solved, but recast only into fine-tunings
between brane and bulk cosmological constants\cite{gibbons}.
However, it has been shown that it is possible to have a flat
solution without fine-tunings of input parameters in the RS II
model with the addition of the bulk three form field\cite{kklcc},
signaling a possible existence of the solution of the cosmological
constant problem.

One more thing to note is that in the RS I model the TeV brane
takes a negative tension while the Planck brane takes a positive
tension, which introduced some cosmological difficulties toward a
smooth transition to the standard big bang cosmology\cite{cosmo}.
This cosmological problem in the RS I model can be solved by
taking into account the radius stabilization\cite{rad}. But it is
worthwhile to search for models with the visible brane taking a
positive tension or branes taking positive tensions only.

In this context, when the higher curvature term is included as a
Gauss-Bonnet invariant in the RS I model, one can find that there
exists another flat solution of the RS type under the condition
that the visible brane should have a positive tension\cite{kklgb}.
However, it is necessary to deal with both the Einstein-Hilbert
term and the Gauss-Bonnet term on the equal footing, but not the
Gauss-Bonnet term as being subdominant compared to the
Einstein-Hilbert term. Other investigations of the higher
curvature gravity concentrated mainly on the Gauss-Bonnet term in
the brane background can be also found in the
literature\cite{gb,CK,kkl,ishwaree}.

On the other hand, we can find more diverse possibilities in
higher dimensional extensions of the RS
model\cite{CK,kkl,ishwaree,ADDK,scalar,kanti}. In fact, in
dimensions higher than $D=5$ the intersecting brane worlds were
considered as a direct generalization of the RS models\cite{ADDK},
but the visible universe does not have a nonzero brane tension,
just regarded as the location where branes with lower codimensions
intersect. The nonzero 3-brane tension in $D>5$ is allowed if the
3-brane is assumed to be made of topological defects supported by
a bulk scalar field\cite{scalar}. In that case, it is shown that
there does not arise fine-tuning condition between brane and bulk
cosmological constants as in the RS model but fine-tuning
conditions appear differently\cite{scalar}.

It is also shown that it is possible to have a nonzero 3-brane
tension in $AdS_6$ intersecting brane worlds by adding the
Gauss-Bonnet term in the bulk action\cite{kkl}. On compactifying
the extra dimensions on an orbifold $T^2/(Z_2\times Z_2)$, the
mass hierarchy can be explained by taking a 3-brane with positive
tension as the visible brane\cite{kkl}. In fact, the $Z_2\times
Z_2$ orbifold symmetry with $Z_2$ acting once on each extra
dimension is nothing but the $Z_2$ orbifold symmetry belonging to
the rotation group around the origin. A higher dimensional($d>6$)
generalization of the intersecting branes in the existence of the
Gauss-Bonnet term have been dealt with in Ref.~\cite{ishwaree}.

Two extra dimension, i.e. 6D, is of particular interest since the
orbifold compactification toward standard-like models in string
theory compactify three two-extra dimensions\cite{orbifold,iknq}.
The $Z_3$ orbifold compactification has been extensively
studied\cite{kim}, which assumed vanishing bulk cosmological
constant. Thus, it is of interest to consider $Z_N$ orbifold
compactification in 6D with negative bulk cosmological constant,
in the hope of obtaining a more general string compactification in
the future. Already, there exists an example that all the SM
matter fields are located at the orbifold fixed
points\cite{kimkim}, the kind of which can be generalized to the
RS I type models.

In this paper, therefore, we extend the orbifold symmetry of the
brane junction in $AdS_6$ to the $Z_N$ case. Then, we also
consider a six-dimensional compactification on another orbifold
$T^2/Z_3$. In that case, there exist three fixed points where
reside 3-branes corresponding to centers of the $Z_3$ symmetric
brane junctions. We show that the 3-brane tensions at these fixed
points are all positive for the Gauss-Bonnet coupling $\alpha>0$.
Then, we can explain the mass hierarchy by regarding two 3-branes,
apart from the 3-brane at the origin, as the visible brane. The
results can be generalized to the orbifold $T^2/Z_N$.

This paper is organized as follows. In Section II, we set up a
general formalism for junctions of semi-infinite 4-branes in
$AdS_6$ in the presence of the Gauss-Bonnet term. In Sec. III we
derive the consistency conditions for the $Z_N$ symmetric brane
junctions and apply the results to the $Z_3$ case. In Sec. IV we
compactify the extra dimensional space with the $Z_3$ symmetric
brane junction on a torus and determine the warp factor and the
3-brane tension located at the fixed points in terms of two
independent 4-brane tensions. Sec. V is a conclusion.

\section{General setup with Gauss-Bonnet term}
The Gauss-Bonnet(GB) term is the only consistent higher curvature
term in the RS models since it does not give rise to higher
derivative terms of the metric beyond the second\cite{kklgb}. In
6D, it was also shown that the theory with the GB term can
accomodate 3-branes with nonzero tensions in the 6D
spacetime\cite{kkl}. Thus, when we include the GB term as the next
leading-order higher curvature interaction on top of the
conventional Einstein-Hilbert term, the 6D bulk action reads
\begin{eqnarray}
S_6=\int d^4x dz_1 dz_2\sqrt{-g}\bigg[ {M^4\over 2}R-\Lambda_b
+\frac{1}{2}\alpha M^2 (R^2 -4R_{MN}R^{MN}+R_{MNPQ}R^{MNPQ})\bigg]
\label{action}
\end{eqnarray}
where $M$ is the six dimensional gravitational constant,
$\Lambda_b$ is the bulk cosmological constant,
$\alpha$ is the effective coupling.

If we assume the metric ansatz as
\begin{eqnarray}
ds^2_6&=&A^2(z_1,z_2)(\eta_{\mu\nu}dx^\mu dx^\nu+dz_1^2+dz_2^2),\label{metric}
\end{eqnarray}
where $(\eta_{\mu\nu})=diag.(-1,+1,+1,+1)$,
it has been shown\cite{kkl} that the bulk solution of the warp factor is
\begin{eqnarray}
A^{-1}(z_1,z_2)=\vec{k}\cdot\vec{z}+c_0\label{sol}
\end{eqnarray}
where $\vec{k}=(k_{z_1},k_{z_2})$, $\vec{z}=(z_1,z_2)$ and $c_0$
is an integration constant. In each 4-brane patch, one $\vec k$ is
defined. Thus, when we consider multi branes, we will denote $N$
such vectors as $\vec k_{(l)}(l=1,2,\cdots N)$. However, the
magnitude of $\vec{k}$ is fixed in terms of the bulk cosmological
constant $\Lambda_b$ and the effective coupling $\alpha$ as
\begin{equation}
k_{z_1}^2+k_{z_2}^2
=\frac{M^2}{12\alpha}\bigg[1\pm
\sqrt{1+\frac{12\alpha\Lambda_b}{5M^6}}\bigg]
\equiv k_\pm^2.\label{bulk}
\end{equation}
When we introduce singular brane sources in the bulk, the boundary
conditions at a 4-brane with tension $\Lambda$ and a 3-brane with
tension $\lambda$ become\cite{kkl},
\begin{equation}
4\bigg(1-\frac{12\alpha k^2_\pm}{M^2}\bigg)
\bigg(\frac{A'}{A^2}\bigg)\Big|^{+}_{-} =-\frac{\Lambda}{M^4}\ \
(4-{\rm brane}) \label{bc4}
\end{equation}
\begin{equation}
\frac{24\alpha}{M^2}\bigg(\frac{A'}{A^2}\bigg)\Big|^{+}_{-}
\bigg(\frac{\dot{A}}{A^2}\bigg)\Big|^{+}_{-} =\frac{\lambda}{M^4}\
\ (3-{\rm brane}), \label{bc3}
\end{equation}
respectively, where the prime for the case of the 4-brane denotes
the derivative with respect to the bulk coordinate normal to the
4-brane and the prime and dot for the case of the 3-brane denote
the derivatives with respect to any set of two orthogonal bulk
coordinates for the case of two orthogonally intersecting
4-branes\cite{kkl}. Note that here $\cdot$ is not a derivative
with respect to $t$.

Let us consider a junction of semi-infinite $L$ 4-branes and one
3-brane residing on the brane junction in the presence of the GB
term as shown in Fig.~1. In order to obtain the solutions of the
Einstein equations in the presence of singular brane sources in
the extra dimensions, we only have to glue the patches of
different $A$'s between 4-branes such that the metric is
continuous at the locations of the branes and the discontinuities
of the derivatives reproduce the energy momentum tensor of the
branes. Note that we assume the same bulk parameters in all
patches and thus $\vec{k}=(k_{z_1},k_{z_2})$ in each patch is
constrained by Eq.~(\ref{bulk}).

Then we can write the warp factor in a compact form for a space
composed of $L$ $AdS$ patches
\begin{equation}
A^{-1}=c_0+\sum^L_{l=1}(\vec{k}_l\cdot\vec{z})\theta(\vec{n}_{l-1}\cdot\vec{z})
\theta(-\vec{n}_l\cdot\vec{z}),\label{warp}
\end{equation}
where $\vec{n}_l=(-{\rm sin}\varphi_l,{\rm cos}\varphi_l)$ is a unit vector
in the $z_1-z_2$ plane normal to the $l^{th}$ 4-brane,
and $\varphi_l$ is the angle between the $l^{th}$ 4-brane and the $z_1$ axis.
We can always set $\varphi_L=0$ up to the overall rotation
of the configuration.

Let us turn to the energy momentum tensor of the configuration of
$L$ $AdS$ patches separated by 4-branes at the junction of which a
3-brane is located as in Fig.~1. We recall that the bulk energy
momentum tensor is assumed to be the same for all bulk spaces as
$T^{bulk}_{MN}=-\Lambda_b g_{MN}$. On the other hand, the energy
momentum tensor for branes is
\begin{equation}
T^{brane}_{MN}=\sum^L_{l=1}T^{4-brane,l}_{MN}+T^{3-brane}_{MN},
\end{equation}
where
\begin{equation}
T^{4-brane,l}_{MN}=\Lambda_l A(z_1,z_2)\delta(\vec{n}\cdot\vec{z})
\left(\begin{array}{rrrrrr}
1& & & & & \\ &-1& & & & \\ & &-1& & & \\ & & &-1& & \\
 & & & &-{\rm cos}^2\varphi_l&-{\rm sin}\varphi_l{\rm cos}\varphi_l \\
 & & & &-{\rm sin}\varphi_l{\rm cos}\varphi_l&-{\rm sin}^2\varphi_l
\end{array}\right),
\end{equation}
and
\begin{equation}
T^{3-brane}_{MN}=\lambda_1\delta(z_1)\delta(z_2)\eta_{\mu\nu}\delta^\mu_M
\delta^\nu_N.
\end{equation}

Note that the warp factors satisfy
\begin{equation}
\frac{A^\prime}{A^2}=-k_{z_1},\ \ \frac{\dot A}{A^2}=-k_{z_2}
\end{equation}
Then, on inspecting
Eqs.~(\ref{bc4}) and (\ref{bc3}), we find the boundary conditions
matching the discontinuities of derivatives with brane
singularities as
\begin{equation}
\gamma\Delta \vec{k}_l
=\gamma(\vec{k}_{l+1}-\vec{k}_l)
=\frac{\Lambda_l}{4M^4}\vec{n}_l,
 \,\,\, (l=1,2,\dots, L), \,\,\,{\rm with}\,\,\,
\vec{k}_{L+1}=\vec{k}_1,\label{junc1}
\end{equation}
\begin{equation}
(\vec{k}_1-\vec{k}_{[L/2]})_{z_1}(\vec{k}_{[L/4]}-\vec{k}_{[3L/4]})_{z_2}
=\frac{\lambda_1}{24\alpha M^2},\label{junc2}
\end{equation}
where
\begin{equation}
\gamma=1-\frac{12\alpha k^2_\pm}{M^2}
\end{equation}
and $[x]$ denotes the largest integer not exceeding $x$. Here we
note that the polygonal integration near the origin of the 3-brane
is assumed in deriving the boundary condition (\ref{junc2}) but it
can be easily shown that it is equivalent to the square
integration as in the case of two orthogonally intersecting
4-branes. After summing up Eq.~(\ref{junc1}), we obtain a
condition for the brane sources,
\begin{equation}
\sum^L_{l=1}\Lambda_l\vec{n}_l=0 \,\,\, {\rm or}\,\,\,
\sum^L_{l=1}\vec{\Lambda}_l=0, \label{zerof}
\end{equation}
where $\vec{\Lambda}_l\equiv(\Lambda_l{\cos}\varphi_l,
\Lambda_l{\sin}\varphi_l)$ is defined to point to the brane
direction with angle $\varphi_l$ with respect to the $L^{\rm th}$
brane.

Consequently, there arise $3L+1$ equations(Eqs. (\ref{bulk}),
(\ref{junc1}), and (\ref{junc2})) with respect to the $3L-1$
parameters($k_{l,z_1},k_{l,z_2}$ and $\varphi_l$)
of the ansatz (\ref{metric}) and (\ref{warp}), which
would give rise to two fine-tuning conditions for consistency.

\section{The $Z_N$ symmetric brane junctions in 6D}
Let us now impose the discrete rotation symmetry $Z_N$ on the
brane junction solutions obtained in the previous section. In this
case, we must investigate the additional requirements arising from
the brane junctions $Z_N$ symmetry. For this purpose, we let the
$AdS$ patches between 4-branes to be equally spaced, i.e.,
$\varphi_l=2\pi l/L$ with $l=1,2,\dots, L$.

For conveniece, we can rewrite the warp factor (\ref{warp})
for the brane junction by using complex numbers as
\begin{equation}
A^{-1}=c_0+\sum^L_{l=1}\frac{1}{2}(\bar{k}_l z+k_l\bar{z})
\theta(\frac{e^{-i\varphi_{l-1}}z-e^{i\varphi_{l-1}}\bar{z}}{2i})
\theta(\frac{e^{-i\varphi_l}z-e^{i\varphi_l}\bar{z}}{-2i})
\end{equation}
where $k_l=k_{l,z_1}+ik_{l,z_2}$ and $z=z_1+iz_2$. In order to
make the brane junction invariant under a $Z_N$ rotation,
$z\rightarrow e^{i(2\pi n/N)}z$ with $n=1,2,\dots, N$, we obtain
the complex number $k_l$ transforms as
\begin{equation}
k'_l=e^{-i(2\pi n/N)}k_l=k_{l'},
\end{equation}
with the angle rotated to
\begin{equation}
\varphi'_l=\varphi_l-\frac{2\pi n}{N}=\varphi_{l'},
\end{equation}
from which $l'=l-nL/N$.
Therefore, we obtain the following consistency conditions for the $Z_N$
symmetric brane junction :
\begin{equation}
k_l=e^{i(2\pi n/N)}k_{l-rn},\,\,\,(l=1,2,\dots,L,\,\,\,n=1,2,\dots,N,\,\,\,
l-rn>0)\label{zn}
\end{equation}
where $r=L/N$ is assumed to be a natural number.

For instance, let us consider the case with the $Z_3$ symmetric
brane junction. We also take the number of $AdS$ patches or
4-branes as\\
\indent Case (A) $L=12$, and\\
\indent Case (B) $L=6$\\
for future use in considering the $T^2/Z_3$ orbifold. We include
3-brane tensions at the brane junctions for both cases.

To begin with, for Case (A) with $L=12$, from Eq.~(\ref{zn}), we
obtain the consistency condition for the $Z_3$ symmetry as
$k_l=e^{i(2\pi n/3)}k_{l-4n}$ with $l=1,2,\dots,L$ with $L=12$ and
$n=1,2,3$, which implies only four independent $AdS$ patches or
4-branes. Then, from Eq.~(\ref{junc1}), the boundary conditions at
the four independent 4-branes ($\Lambda_1$, $\Lambda_2$,
$\Lambda_3$ and $\Lambda_{12}$) become
\begin{eqnarray}
\gamma(\vec{k}_2-\vec{k}_1)&=&\frac{\Lambda_1}{8M^4}(-1,\sqrt{3}),\label{4b1}\\
\gamma(\vec{k}_3-\vec{k}_2)&=&\frac{\Lambda_2}{8M^4}(-\sqrt{3},1), \label{4b2}\\
\gamma(\vec{k}_4-\vec{k}_3)&=&\frac{\Lambda_3}{8M^4}(-2,0),\label{4b3}\\
\gamma(\vec{k}_1-\vec{k}_{12})&=&\frac{\Lambda_{12}}{8M^4}(0,2),\label{4b4}
\end{eqnarray}
where $k_{12}$ is related to $k_4$ by the $Z_3$ symmetric
condition, Eq.~(\ref{zn}), as
\begin{equation}
k_{12}=e^{i(4\pi/3)}k_4.
\end{equation}
Then, we can rewrite the $k_l$'s in terms of $k_1$ and the 4-brane tensions as
\begin{eqnarray}
\vec{k}_2&=&\vec{k}_1+\frac{\Lambda_1}{8\gamma M^4}(-1,\sqrt{3}),\label{b1}\\
\vec{k}_3&=&\vec{k}_1+\frac{1}{8\gamma M^4}
(-\Lambda_1-\sqrt{3}\Lambda_2,\sqrt{3}\Lambda_1+\Lambda_2),\label{b2}\\
\vec{k}_4&=&\vec{k}_1+\frac{1}{8\gamma M^4}
(-\Lambda_1-\sqrt{3}\Lambda_2-2\Lambda_3,\sqrt{3}\Lambda_1+\Lambda_2),
\label{b3}\\
\vec{k}_{12}&=&\vec{k}_1+\frac{\Lambda_{12}}{8\gamma M^4}(0,-2),\label{b4}
\end{eqnarray}
where $k_{12}$ has additional relations as
\begin{eqnarray}
k_{12,z_1}&=&\frac{1}{2}(-k_{4,z_1}+\sqrt{3}k_{4,z_2})\nonumber\\
&=&\frac{1}{2}(-k_{1,z_1}+\sqrt{3}k_{1,z_2})
+\frac{1}{8\gamma M^4}(2\Lambda_1+\sqrt{3}\Lambda_2+\Lambda_3),\label{tw1} \\
k_{12,z_2}&=&\frac{1}{2}(-\sqrt{3}k_{4,z_1}-k_{4,z_2})\nonumber\\
&=&\frac{1}{2}(-\sqrt{3}k_{1,z_1}-k_{1,z_2})
+\frac{1}{8\gamma M^4}(\Lambda_2+\sqrt{3}\Lambda_3).\label{tw2}
\end{eqnarray}
In view of Eq.~(\ref{bulk}), we have a universal
$\vec{k}^2_l=\vec{k}^2_1=k^2_\pm$ for all $l$. Thus, taking the
square of each of Eqs.~(\ref{b1})-(\ref{b4}) on both sides gives,
respectively,
\begin{equation}
\Lambda_1\bigg(-k_{1,z_1}+\sqrt{3}k_{1,z_2}
+\frac{\Lambda_1}{4\gamma M^4}\bigg)=0,\label{cond1}
\end{equation}
\begin{equation}
\Lambda_2\bigg(-\sqrt{3}k_{1,z_1}+k_{1,z_2}
+\frac{1}{4\gamma M^4}(\sqrt{3}\Lambda_1+\Lambda_2)\bigg)=0, \label{cond2}
\end{equation}
\begin{equation}
\Lambda_3\bigg(-k_{1,z_1}
+\frac{1}{8\gamma M^4}(\Lambda_3+\Lambda_1+\sqrt{3}\Lambda_2)\bigg)=0,
\label{cond3}
\end{equation}
\begin{equation}
\Lambda_{12}\bigg(-k_{1,z_2}+\frac{\Lambda_{12}}{8\gamma M^4}\bigg)=0.
\label{cond4}
\end{equation}
Therefore, there are three cases consistent with the $Z_3$ symmetry
\begin{eqnarray}
{\rm (i)}~~&:&~~\Lambda_3=\Lambda_1\neq 0,~~\Lambda_{12}=\Lambda_2\neq 0,
\label{cc1}\\
{\rm (ii)}~~&:&~~\Lambda_3=\Lambda_1=0,~~\Lambda_{12}\neq 0,
~~\Lambda_2\neq 0,\label{cc2}\\
{\rm (iii)}~~&:&~~\Lambda_{12}=\Lambda_2=0,~~\Lambda_{3}\neq 0,
~~\Lambda_1\neq 0.\label{cc3}
\end{eqnarray}
Here we keep both the cases (ii) and (iii), which will be useful
for the $T^2/Z_3$ orbifold in the next section even though
those configurations themselves are equivalent to each other
up to a rotation.
Then, we determine $k_1$ by solving the equations (\ref{cond1})-(\ref{cond4})
consistently with Eqs.~(\ref{b4})-(\ref{tw2}) for each case as
\begin{eqnarray}
{\rm (i)}~~&:&~~k_{1,z_1}=\frac{1}{8\gamma M^4}
(2\Lambda_1+\sqrt{3}\Lambda_2),~~
k_{1,z_2}=\frac{\Lambda_2}{8\gamma M^4},\label{c1}\\
{\rm (ii)}~~&:&~~k_{1,z_1}=\frac{1}{8\sqrt{3}\gamma M^4}(\Lambda_{12}
+2\Lambda_{2}), ~~k_{1,z_2}=\frac{\Lambda_{12}}{8\gamma M^4}\label{c2}\\
{\rm (iii)}~~&:&~~k_{1,z_1}=\frac{1}{8\gamma M^4}(\Lambda_1+\Lambda_3),~~
k_{1,z_2}=\frac{1}{8\sqrt{3}\gamma M^4}(\Lambda_3-\Lambda_1).\label{c3}
\end{eqnarray}
On the other hand, from Eq.~(\ref{junc2}), the boundary condition
at the 3-brane becomes
\begin{equation}
(k_1-k_6)_{z_1}(k_4-k_9)_{z_2}=\frac{\lambda_1}{24\alpha M^2}.
\end{equation}
Since $k_6=e^{i(2\pi/3)}k_2$ and $k_9=e^{i(4\pi/3)}k_1$, using
Eq.~(\ref{b1}), (\ref{b3}) and (\ref{c1})-(\ref{c3}), we obtain a fine-tuning
relation involving both 4-brane and 3-brane tensions for each case as
\begin{eqnarray}
{\rm (i)}~~&:&~~\lambda_1=\frac{3\alpha}{2\gamma^2 M^6}(2\Lambda_1
+\sqrt{3}\Lambda_2)(\sqrt{3}\Lambda_1+2\Lambda_2),\label{tunec1}\\
{\rm (ii)}~~&:&~~\lambda_1=\frac{3\sqrt{3}\alpha}{4\gamma^2 M^6}
(\Lambda_{12}+\Lambda_2)^2,\label{tunec2}\\
{\rm (iii)}~&:&~~\lambda_1=\frac{3\sqrt{3}\alpha}{4\gamma^2 M^6}
(\Lambda_1+\Lambda_3)^2. \label{tunec3}
\end{eqnarray}
Thus we find that
the 3-brane located at the junction of 4-branes
with $L=12$ for any case of (\ref{cc1})-(\ref{cc3})
should have a {\it positive tension} for $\alpha>0$
(from the positivity of $k_{1,z_1}$ and $k_{1,z_2}$
in Eq.~(\ref{c3}) for the case (i)).

For Case (B) with $L=6$, we also obtain the $Z_3$ symmetric
condition as $k'_l=e^{i(2\pi n/3)}k'_{l-2n}$ with $l=1,2,\dots, 6$
and $n=1,2,3$, which implies only two independent $AdS$ patches or
4-branes ($V_1$ and $V_6$). Therefore, we do not need to do
another calculation for that since Case (B) with $L=6$ corresponds
to Case (A)--(ii) when a nonzero 4-brane tension is taken to
coincide with the positive half of the $z_1$ axis. We also
determine $k'_1$ and the 3-brane tension located at the brane
junction, respectively,
\begin{equation}
k'_{1,z_1}=\frac{1}{8\sqrt{3}\gamma M^4}(V_6+2V_1),
~~k'_{1,z_2}=\frac{V_6}{8\gamma M^4},\label{caseb1}
\end{equation}
\begin{equation}
\lambda_1=\frac{3\sqrt{3}\alpha}{4\gamma^2 M^6}
(V_1+V_6)^2.\label{caseb2}
\end{equation}
Here we also observe that the 3-brane located at the junction of
4-branes with $L=6$ should a {\it positive tension} for
$\alpha>0$.

\section{mass hierarchy with the orbifold $T^2/Z_3$}
In the previous section, we dealt with the $Z_N$ symmetric brane junction
solutions for the non-compact extra dimensions.
We can also consider the case with compact extra dimensions
of the $Z_N$ invariance through taking into account the $Z_N$ symmetric brane
junction solutions. In this section, for the purpose of using
the exact solutions for the $Z_3$ symmetric brane junction solutions
obtained in the previous section,
we take the orbifold $T^2/Z_3$ as a geometry of extra dimensions.
The case with flat extra dimensions was investigated in the orbifold
constructions of superstring compactifications\cite{orbifold}.

When we consider a torus with $Z_3$ invariance, we need to identify the extra
dimensions in the following way :
\begin{eqnarray}
T^2/Z_3:~~z\approx z+a(n_1+n_2 e^{i(2\pi/3)})
\end{eqnarray}
where $z=z_1+i z_2$, $a$ is the size of each extra dimension and
$(n_1,n_2)$ is an integer-valued lattice vector. Then, there
appear three fixed points with $z=\frac{ka}{\sqrt{3}}e^{i\pi/6}$
with $k=0,1,2$. Identification of the orbifold $T^2/Z_3$ is shown
in Fig.~2. Essentially, we find that the extra dimension space is
composed of one brane junction (around {\Large $\bullet$}) with
$L=12$ $AdS$ patches (Case (A)) and two brane junctions (around
$\blacksquare$ and around $\blacktriangle$) with $L=6$ $AdS$
patches (Case (B)), each of which is shown to be $Z_3$ symmetric
around its center corresponding to one of fixed points. [The
similar things happen for other orbifolds : two $Z_4$ symmetric
brane junctions with $L=8$ and one $Z_4$ symmetric brane junction
with $L=4$ for an $T^2/Z_4$ orbifold while one $Z_6$ symmetric
brane junction with $L=12$ and one $Z_3$ symmetric brane junction
with $L=6$ for an $T^2/Z_6$ orbifold.] Thus, we can easily
identify 4-brane and 3-brane tensions in the fundamental region by
using results on the $Z_3$ brane junction solutions obtained in
the previous section. We denote independent 4-brane tensions with
$(\Lambda_1,\Lambda_2)$ around {\Large $\bullet$}, $(V'_1,V'_6)$
around $\blacksquare$ and
$(V^{\prime\prime}_1,V^{\prime\prime}_6)$ around $\blacktriangle$.
And hereafter we use a simple notation $(k_1,k_2)$ for
$\vec{k}_1$, for example, in Eqs.~(\ref{c1})-(\ref{c3}).

However, since we chose the coordinate system such that a 4-brane
coincides with the positive half of one of bulk coordinates, in
the common region of Cases (A) and (B), we need to write $k'_l$
and $k^{\prime\prime}_l$ defined in Case (B) in terms of $k_l$
defined in Case (A). To begin with, using the solution of Case
(A), we can find the warp factor in the regions of Case (B) by the
orbifold symmetry. Particularly, when the warp factor in the
region (I) of Fig.~2 is given by the solution of Case (A), the
warp factor in the regions (II) and (III) of Fig.~2 are also
written in bulk coordinates adopted for Cases (A) and (B). The
results read
\begin{eqnarray}
{\rm (I)}:~~A^{-1}&=&k_1z_1+k_2z_2+1,\label{rg1}\\
{\rm (II)}:~~A^{-1}&=&-k_1z_1+k_2 z_2+k_1a+1
=k'_1z'_1+k'_2z'_2+c'_0, \label{rg2}\\
{\rm (III)}:~~A^{-1}&=&k_1z_1-k_2 z_2+1
=k^{\prime\prime}_1 z^{\prime\prime}_1
+k^{\prime\prime}_2 z^{\prime\prime}_2
+c^{\prime\prime}_0 \label{rg3}~~{\rm for}~~{\rm (A)-(i),(ii)},\\
A^{-1}&=&k_1z_1+k_2 z_2+1
=k^{\prime\prime}_1 z^{\prime\prime}_1
+k^{\prime\prime}_2 z^{\prime\prime}_2
+c^{\prime\prime}_0 \label{rg3a}~~{\rm for}~~{\rm (A)-(iii)},
\end{eqnarray}
where $-z'_1$ and $z^{\prime\prime}_1$ axes are chosen to coincide
with $z_1=a/2$. Since $z_1=z'_2=-z^{\prime\prime}_2$ and
$z_2=-z'_1=z^{\prime\prime}_1$, we obtain $k'_1=\pm
k^{\prime\prime}_1=-k_2$ (+ for Cases (A)--(i),(ii), and -- for
Case (A)--(iii)) and $k'_2=k^{\prime\prime}_2=-k_1$. Thus we can
regard the warp factor in the extra dimensions as being determined
only by $(k_1,k_2)$, which are given by 4-brane tensions belonging
to Case (A) in view of Eqs.~(\ref{c1})-(\ref{c3}) as follows
\begin{eqnarray}
{\rm (i)}~~&:&~~k_1=\frac{1}{8\gamma M^4}
(2\Lambda_1+\sqrt{3}\Lambda_2),~~
k_2=\frac{\Lambda_2}{8\gamma M^4},\label{c1a}\\
{\rm (ii)}~~&:&~~k_1=\frac{1}{8\sqrt{3}\gamma M^4}(\Lambda_{12}
+2\Lambda_{2}), ~~k_2=\frac{\Lambda_{12}}{8\gamma M^4},\label{c2a}\\
{\rm (iii)}~~&:&~~k_1=\frac{1}{8\gamma M^4}(\Lambda_1+\Lambda_3),~~
k_2=\frac{1}{8\sqrt{3}\gamma M^4}(\Lambda_3-\Lambda_1).\label{c3a}
\end{eqnarray}
Therefore, it is easy to show that the 4-brane tensions belonging to Case
(B) do not remain independent from Eq.~(\ref{caseb1}) but they should
be related to the 4-brane tensions belonging to Case (A) as
\begin{eqnarray}
{\rm (i)}~~&:&~~V'_6=V^{\prime\prime}_6=-(2\Lambda_1+\sqrt{3}\Lambda_2),
~~V'_1=V^{\prime\prime}_1=\Lambda_1, \label{4tension1}\\
{\rm (ii)}~~&:&~~V'_6=V^{\prime\prime}_6=-\frac{1}{\sqrt{3}}
(\Lambda_{12}+2\Lambda_{2}),
~~V'_1=V^{\prime\prime}_1=\frac{1}{\sqrt{3}}(\Lambda_2-\Lambda_{12})=0,
\label{4tension2}\\
{\rm (iii)}~~&:&~~V'_6=V^{\prime\prime}_6=-(\Lambda_1+\Lambda_3),
~~V'_1=\Lambda_1,~~V^{\prime\prime}_1=\Lambda_3,\label{4tension3}
\end{eqnarray}
where we used $V'_1=\Lambda_1=0$ in Eq.~(\ref{4tension2})
from the $Z_3$ identification in Fig.~2.
On the other hand, the 3-brane tensions at the three fixed points are
determined in terms of 4-brane tensions from
Eqs.~(\ref{tunec1}) and (\ref{caseb2}) for Case (A)--(i) to begin with
\begin{eqnarray}
{\rm (i)}~~:~~\lambda_1&=& \frac{3\alpha}{2\gamma^2 M^6}
(2\Lambda_1+\sqrt{3}\Lambda_2) (\sqrt{3}\Lambda_1+2\Lambda_2),\\
\lambda_2&=&\frac{3\sqrt{3}\alpha}{4\gamma^2 M^6}(V'_1+V'_6)^2
=\frac{3\sqrt{3}\alpha}{4\gamma^2 M^6}(\Lambda_1+\sqrt{3}\Lambda_2)^2,\\
\lambda_3&=&\frac{3\sqrt{3}\alpha}{4\gamma^2 M^6}
(V^{\prime\prime}_1+V^{\prime\prime}_6)^2=\lambda_2,
\end{eqnarray}
and for Cases (A)--(ii) and (iii) as
\begin{eqnarray}
{\rm (ii)}~~:~~\lambda_1&=&\frac{3\sqrt{3}\alpha}{4\gamma^2 M^6}
(\Lambda_{12}+\Lambda_2)^2,
~~\lambda_2=\frac{9\sqrt{3}\alpha}{4\gamma^2 M^6}\Lambda_2^2=\lambda_3,\\
{\rm (iii)}~~:~~\lambda_1&=&\frac{3\sqrt{3}\alpha}{4\gamma^2 M^6}
(\Lambda_1+\Lambda_3)^2,
~~\lambda_2=\frac{3\sqrt{3}\alpha}{4\gamma^2 M^6}\Lambda_3^2,
~~\lambda_3=\frac{3\sqrt{3}\alpha}{4\gamma^2 M^6}\Lambda_1^2.
\end{eqnarray}

Moreover, in view of Eqs.~(\ref{rg1})-(\ref{rg3a}),
we also need to have another 3-brane tension at a
non-fixed point $(z_1,z_2)=(a/2,0)$ (and at its $Z_3$ transformed
points) only for Cases (A)--(i) and (ii)
\begin{eqnarray}
{\rm (i)}~~:~~\lambda_4&=&-\frac{3\alpha}{2\gamma^2 M^6}\Lambda_2
(2\Lambda_1+\sqrt{3}\Lambda_2), \\
{\rm (ii)}~~:~~\lambda_4&=&-\frac{3\sqrt{3}\alpha}{2\gamma^2 M^6}\Lambda_2^2.
\end{eqnarray}
Here we find that the 3-brane tensions at the fixed points are all
positive for $\alpha>0$ but all negative for $\alpha<0$ (from the
positiveness of $k_1$ and $k_2$ in Eq.~(\ref{c1a}) for case (i)).
In addition, there should exist an extra 3-brane(the
$\lambda_4$-brane) which is not located at the fixed points only
for Cases (A)--(i) and (ii) but there does not arise such an
additional 3-brane for Case (A)--(iii). Note that in order to
explain the large mass hierarchy with extra dimensions
compactified on the $T^2/Z_3$ orbifold, we may take the
$\lambda_2$ or $\lambda_3$ branes as the visible brane with
positive tension for $\alpha>0$ while the $\lambda_1$ brane can be
considered as the hidden brane at the Planck scale.

For our purpose of showing the generation of the large mass hierarchy,
let us rewrite the metric as
\begin{eqnarray}
ds^2_6&=&A^2(z_1,z_2)(\eta_{\mu\nu}dx^\mu dx^\nu+dz_1^2+dz_2^2)\nonumber\\
&=&A^2(y_1,y_2)\eta_{\mu\nu}dx^\mu dx^\nu+B^2(y_1,y_2)dy_1^2
+C^2(y_1,y_2)dy_2^2
\end{eqnarray}
by the following bulk coordinate transformations:
\begin{eqnarray}
dz_1=\frac{B}{A}dy_1,\,\,dz_2=\frac{C}{A}dy_2.
\end{eqnarray}
For example, $k_1z_1=e^{k_1y_1}-1$, $k_2z_2=e^{k_2y_2}-1$ for
$A^{-1}=k_1z_1+k_2z_2+1$. Then, we can have the warp factor in the
new coordinate: $A=(e^{k_1y_1}+e^{k_2y_2}-1)^{-1}$,
$B=e^{k_1y_1}A$ and $C=e^{k_2y_2}A$. In this new coordinate, let
us consider the action for the Higgs scalar field at the
$\lambda_2$ brane connecting to the patch (I) for all cases of (A)
\begin{eqnarray}
S_{vis}\supset\int d^4 x \sqrt{-g^{(vis)}}\bigg[\bar{g}^{\mu\nu}
\partial_\mu H\partial_\nu H-(H^2-m_0^2)^2\bigg],\nonumber \\
=\int d^4 x \sqrt{-g^{(4)}}A^4\bigg[A^{-2}(\partial
H)^2-(H^2-m_0^2)^2\bigg]
\end{eqnarray}
where $m_0$ is of order the Planck scale. Then, redefining the
scalar field as $\tilde{H}=AH$ gives
\begin{eqnarray}
\int d^4 x\sqrt{-g^{(4)}}\bigg[(\partial \tilde{H})^2
-(\tilde{H}^2-m^2_2)^2\bigg]
\end{eqnarray}
where the Higgs mass parameter on the $\lambda_2$ brane is given by
\begin{eqnarray}
m_2=Am_0=(e^{k_1 b_1}+e^{k_2 b_2}-1)^{-1}m_0
\end{eqnarray}
with
\begin{eqnarray}
b_1&=&\frac{1}{k_1}{\rm log}\Big(\frac{1}{2}k_1 a+1\Big),\\
b_2&=&\frac{1}{k_2}{\rm log}\Big(\frac{\sqrt{3}}{6}k_2 a+1\Big).
\end{eqnarray}
Similarly, we obtain the effective mass scale on the $\lambda_3$ brane
connecting to the patch (III)
\begin{eqnarray}
{\rm (i),(ii)}~~&:&~~m_3=(e^{k_1 b_1}+e^{k_2 b_2}-1)^{-1}m_0=m_2, \\
{\rm (iii)}~~&:&~~m_3=(e^{k_1 b_1}+e^{-k_2 b_2}-1)^{-1}m_0.
\end{eqnarray}
As a result, we can obtain two weak scale branes with positive
tensions located at two fixed points of the $T^2/Z_3$ orbifold.
Therefore, the hierarchy problem related to the Higgs mass can be
explained by regarding one of two weak scale branes as the visible
brane while the $\lambda_1$ brane with the Planck scale at the
origin is taken as the hidden brane. These two scales can be used
to solve the hierarchy problem as in the RS I case. Or we can use
it for the third generation and the first two generations mass
hierarchy problem(the third generation matter located at the
$\lambda_1$-brane) if the gauge hierarchy problem is solved by
supersymmetry. Except for the Case (iii), the $\lambda_4$-brane
can be used to locate the untwisted matter fields if it is
required to put matter fields at the 3-branes. For Case (iii),
however, it is not required to put a $\lambda_4$-brane and hence
the string models with matter arising only from twisted sectors
belong to this category\cite{kimkim}.

As in the case of $T^2/Z_2$ discussed in Ref.\cite{kkl} and
$T^2/Z_3$ orbifolds considered in this section, it turns out that
other possible orbifolds with discrete rotation symmetry such as
$T^2/Z_4$ and $T^2/Z_6$ can give rise to similar candidates for
the visible brane with positive tension to solve the gauge
hierarchy or flavor hierarchy problem. However, only for the
$T^2/Z_3$ case we can take all the 3-brane tensions at the fixed
points to be positive.

\section{Conclusion}
We formulated a $Z_N$ symmetric RS models in $AdS_6$ with the
Gauss-Bonnet term. Firstly, we considered a junction of
semi-infinite 4-branes with the $Z_N$ discrete rotation symmetry
in $AdS_6$ with a Gauss-Bonnet term. The Gauss-Bonnet term gives
rise to more divergent terms than the case with 4-branes only to
accomodate a nonzero 3-brane tension at the junction of 4-branes.
Then the extra dimensional space with the $Z_N$ discrete symmetry
due to the brane junction is compactified on a torus, we find that
the bulk space is separated into a number of distinguishable
regions(for example a, b, c, and d of Fig. 2 for the $Z_3$
orbifold) by the brane junctions with discrete symmetries around
their centers. Our particular interest is on the $Z_3$ orbifold
which renders a possible solution of the three family
problem~\cite{orbifold}. Hence, we identified the 3-brane tensions
at the fixed points of an $T^2/Z_3$ orbifolds by using the results
on the $Z_3$ symmetric brane junctions. As a result, the three
3-brane tensions can be positive with a positive Gauss-Bonnet
coupling ($\alpha>0$). We also observed that the gauge hierarchy
can be explained by regarding one of two weak scale 3-branes as
the visible brane. The results can be generalized to the case of
the $T^2/Z_N$ orbifold but it is the case only for the $T^2/Z_3$
orbifold that we can take all the 3-brane tensions at the fixed
points to be positive, circumventing the cosmological problem of
the negative visible-brane tension in the RS model.

\acknowledgments We thank Bumseok Kyae for helpful inputs when we
started this work together at the initial stage. This work is
supported in part by the BK21 program of Ministry of Education,
the KOSEF Sundo Grant, and by the Center for High Energy
Physics(CHEP), Kyungpook National University.

\vskip 0.3cm
\begin{figure}[b]
\centering
\centerline{\epsfig{file=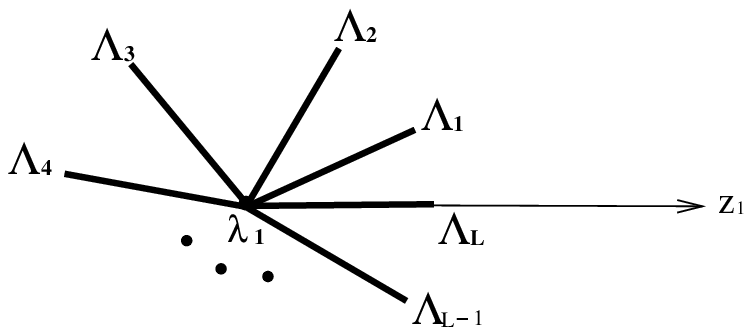,width=90mm}}
\end{figure}
\centerline{ Fig.~1.\ \it A junction of semi-infinite 4-branes
where a 3-brane resides. }
\vskip 0.3cm

\vskip 0.3cm
\begin{figure}[b]
\centering
\centerline{\epsfig{file=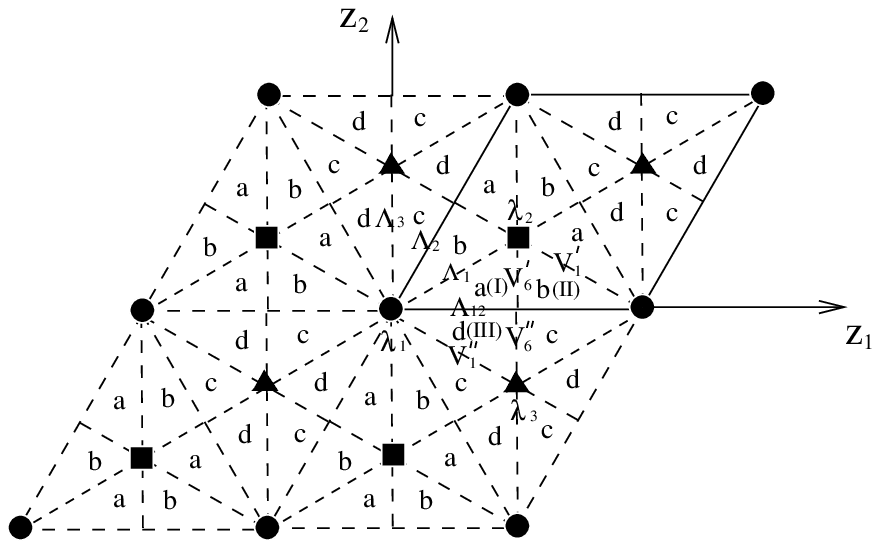,width=130mm}}
\end{figure}
\centerline{ Fig.~2.\ \it The $T^2/Z_3$ orbifold. We distinguish
the regions of the torus under $Z_3$ symmetry}\vskip -0.2cm
\centerline{\it as $a$, $b$, $c$, and $d$. There are three fixed
points depicted by {\Large $\bullet$}, $\blacksquare$ and
$\blacktriangle$.}


\begin{references}

\def\apj#1#2#3{Astrophys.\ J.\ {\bf #1}, #2 (#3)}
\def\ijmp#1#2#3{Int.\ J.\ Mod.\ Phys.\ {\bf #1}, #2 (#3)}
\def\mpl#1#2#3{Mod.\ Phys.\ Lett.\ {\bf #1}, #2 (#3)}
\def\nat#1#2#3{Nature\ {\bf #1}, #2 (#3)}
\def\npb#1#2#3{Nucl.\ Phys.\ {\bf B#1}, #2 (#3)}
\def\plb#1#2#3{Phys.\ Lett.\ {\bf B#1}, #2 (#3)}
\def\prd#1#2#3{Phys.\ Rev.\ {\bf D#1}, #2 (#3)}
\def\prl#1#2#3{Phys.\ Rev.\ Lett.\ {\bf #1}, #2 (#3)}
\def\prt#1#2#3{Phys.\ Rep.\ {\bf #1}, #2 (#3)}
\def\sjnp#1#2#3{Sov.\ J.\ Nucl.\ Phys.\ {\bf #1}, #2 (#3)}
\def\zp#1#2#3{Z.\ Phys.\ {\bf #1}, #2 (#3)}
\def\jhep#1#2#3{JHEP\ {\bf #1}, #2 (#3)}

\bibitem{ADD} N. Arkani-Hamed, S. Dimopoulos and G. Dvali, Phys. Lett.
{\bf B429}, 263 (1998); I. Antoniadis, N. Arkani-Hamed, S. Dimopoulos, and
G. Dvali, Phys. Lett. {\bf B436}, 257 (1998).

\bibitem{rs1} L. Randall and R. Sundrum, \prl{83}{3370}{1999}
[hep-th/9905221].

\bibitem{stab} W. D. Goldbeger and M. B. Wise, \prl{83}{49}{1999}.

\bibitem{rs2} L. Randall and R. Sundrum, \prl{83}{4690}{1999}
[hep-th/9906064].

\bibitem{gravity} J. Garriga and T. Tanaka, \prl{84}{2778}{2000}
[hep-th/9911055]; S. B. Giddings, E. Katz and L. Randall,
\jhep{0003}{023}{2000} [hep-th/0002091]; J. E. Kim and H. M. Lee,
\npb{602}{346}{2001} [hep-th/0010093].

\bibitem{gibbons} G. Gibbons, R. Kallosh abd A. Linde, \jhep{0101}{022}{2001}
[hep-th/0011225].

\bibitem{kklcc} J. E. Kim, B. Kyae and H. M. Lee, \prl{86}{4223}{2001}
[hep-th/0011118] and hep-th/0101027.

\bibitem{cosmo} C. Csaki, M. Graesser, C. Kolda and J. Terning,
\plb{462}{34}{1999} [hep-ph/9906513]; J. M. Cline, C. Grojean and
G. Servant, \prl{83}{4245}{1999} [hep-ph/9906523].

\bibitem{rad} C. Csaki, M. Graesser, L. Randall and J. Terning,
\prd{62}{045015}{2000} [hep-ph/9911406]; H. B. Kim,
\plb{478}{285}{2000} [hep-th/0001209].

\bibitem{kklgb} J. E. Kim, B. Kyae and H. M. Lee,
\prd{62}{045013}{2000} [hep-ph/9912344] and \npb{582}{296}{2000}
[hep-th/0004005].

\bibitem{gb}
I. Low, A. Zee, \npb{585}{395}{2000} [hep-th/0004124]; S. Nojiri
and S. D. Odintsov, \jhep{0007}{049}{2000} [hep-th/0006232]; N. E.
Mavromatos and J. Rizos, \prd{62}{124004}{2000} [hep-th/0008074];
I.P. Neupane, \jhep{0009}{040}{2000} [hep-th/0008191]; O.
Corradini and Z. Kakushadze, \plb{494}{302}{2000}
[hep-th/0009022]; K. A. Meissner and M. Olechowski,
\prl{86}{3708}{2001} [hep-th/0009122]; H. Collins and B. Holdom,
\prd{63}{084020}{2001} [hep-th/0009127]; M. Giovannini,
\prd{63}{064011}{2001} [hep-th/0011153].

\bibitem{CK} O. Corradini and Z. Kakushadze, \plb{506}{167}{2001}
[hep-th/0103031].

\bibitem{kkl} J. E. Kim, B. Kyae and H. M. Lee, \prd{64}{065011}{2001}
[hep-th/0104150].

\bibitem{ishwaree} I. P. Neupane, hep-th/0106100.

\bibitem{ADDK} N. Arkani-Hamed, S. Dimopoulos, G. Dvali and N. Kaloper,
\prl{84}{586}{2000}; C. Csaki and Y. Shifman,
\prd{61}{024008}{2000}; J. M. Cline, C. Grojean and G. Servant,
\plb{472}{302}{2000}.

\bibitem{scalar} A. G. Cohen and D. B. Kaplan,
\plb{470}{52}{1999}; R. Gregory, \prl{84}{2564}{2000}; I.
Olasagasti and A. Vilenkin, \prd{62}{044014}{2000}; T. Gherghetta
and M. Shaposhnikov, \prl{85}{240}{2000} [hep-th/0004014].


\bibitem{kanti} P. Kanti, R. Madden and K. A. Olive, \prd{64}{044021}{2001}
[hep-th/0104177]; F. Leblond, R. C. Meyers and D. J. Winters,
\jhep{0107}{031}{2001} [hep-th/0106140]; I. I. Kogan, S.
Mouslopoulos, A. Papazoglou and G. G. Ross, hep-th/0107086.

\bibitem{orbifold} L. Dixon, J. Harvey, C. Vafa and E. Witten,
\npb{261}{651}{1985} and \npb{274}{285}{1986}.

\bibitem{iknq} L. Ibanez, J. E. Kim, H. P. Nilles, and F. Quevedo,
\plb{191}{282}{1987}.

\bibitem{kim} L. E. Ibanez, J. Mas, H. P. Nilles
and F. Quevedo, \npb{301}{157}{1988}; J. E. Kim,
\plb{207}{434}{1988}; J. A. Casas and C. Munoz,
\plb{214}{63}{1988}; E. J. Chun, J. E. Kim and H. P. Nilles,
\npb{370}{105}{1992}.

\bibitem{kimkim} H. B. Kim and J. E. Kim, \plb{300}{343}{1993}.


\end{references}
\end{document}